\documentclass[runningheads]{llncs}
\usepackage{booktabs} % For formal tables
\usepackage{xspace}
\usepackage{amssymb}
\usepackage{amsmath}
\usepackage{makecell}
\usepackage{booktabs} % For formal tables
\usepackage{todonotes}
\usepackage{customizedmacros}
\usepackage{workinprogressmacros}  
\usepackage{linkeddatamacros}
\usepackage{graphicx}
\usepackage{epstopdf}
\usepackage{url}
\usepackage{algorithm}
\usepackage[noend]{algpseudocode}
\usepackage{soul}
\usepackage{subfig}
\usepackage{multirow}
\usepackage{multicol}
\usepackage{fancyvrb}
\usepackage{wrapfig}
\usepackage{setspace}
\usepackage{stackengine}
\begin{document}
\title{How to make latent factors interpretable by feeding Factorization machines with knowledge graphs}

\author{Vito Walter Anelli\inst{1} \and
	Tommaso Di Noia\inst{1} \and
	Eugenio Di Sciascio\inst{1}\and
	Azzurra Ragone\inst{2}\and
	Joseph Trotta\inst{1}\thanks{Authors are listed in alphabetical order. Contact author: V.W. Anelli (\texttt{vitowalter.anelli@poliba.it})}}
\institute{Polytechnic University of Bari, Bari, Italy \and
	Independent researcher, Milan, Italy
	\email{\{name.surname\}@poliba.it,azzurraragone@gmail.com}}
\titlerunning{How to make latent factors interpretable by feeding FMs with KGs}
%\author{No name}
%
\authorrunning{V.W. Anelli et al.}
%\institute{Princeton University, Princeton NJ 08544, USA \and
%Springer Heidelberg, Tiergartenstr. 17, 69121 Heidelberg, Germany
\maketitle              % typeset the header of the contribution
\begin{abstract}
	
Model-based approaches to recommendation can recommend items with a very high level of accuracy. Unfortunately, even when the model embeds content-based information, if we move to a latent space we miss references to the actual semantics of recommended items. Consequently, this makes non-trivial the interpretation of a recommendation process.
In this paper, we show how to initialize latent factors in Factorization Machines by using semantic features coming from a knowledge graph in order to train an interpretable model. With our model, semantic features are injected into the learning process to retain the original informativeness of the items available in the dataset. The accuracy and effectiveness of the trained model have been tested using two well-known recommender systems datasets. By relying on the information encoded in the original knowledge graph, we have also evaluated the semantic accuracy and robustness for the knowledge-aware interpretability of the final model. 	

\end{abstract}
\section{Introduction}\label{sec:introduction}
Transparency and interpretability of predictive models are gaining momentum since they been recognized as a key element in the next generation of recommendation algorithms.
Interpretability may increase user awareness in the decision-making process and lead to fast (efficiency), conscious and right (effectiveness) decisions. 
When equipped with interpretability of recommendation results, a system ceases to be just a black-box \cite{DBLP:conf/chi/SinhaS02,DBLP:reference/rsh/TintarevM11,DBLP:conf/recsys/Zanker12} and users are more willing to extensively exploit the predictions  \cite{DBLP:conf/icde/TintarevM07,DBLP:conf/cscw/HerlockerKR00}. 
Indeed, transparency increases their trust \cite{DBLP:journals/toit/FalconeSC15} (also exploiting specific semantic structures \cite{drawel2018specification}), and satisfaction in using the system. 
Among interpretable models for Recommender Systems (RS), we may distinguish between those based on Content-based (CB) approaches and those based on Collaborative filtering (CF) ones.
CB algorithms provide recommendations by exploiting the available content and matching it with a user profile \cite{DBLP:conf/adaptive/PazzaniB07,DBLP:journals/umuai/CramerERSRSAW08}. 
The use of content features makes the model interpretable even though attention has to be paid since a CB approach ``\textit{lacks serendipity and requires extensive manual efforts to match the user interests to content profiles}'' \cite{DBLP:journals/corr/abs-1804-11192}. 
On the other hand, the interpretation of CF results will inevitably reflect the approach adopted by the algorithm.  For instance, an item-based and a user-based recommendation could be interpreted, respectively, as "\textit{other users who have experienced A have experienced B}" or "\textit{similar users have experienced B}".
Unfortunately, things change when we adopt more powerful and accurate Deep Learning \cite{DBLP:conf/uic/ChakrabortyTRHA17} or model-based algorithms and techniques for the computation of a recommendation list. Such approaches project items and users in a new vector space of latent features \cite{DBLP:journals/computer/KorenBV09} thus making the final result not directly interpretable. 
In the last years, many approaches have been proposed that take advantage of side information to enhance the performance of latent factor models. Side information can refer to items as well as users \cite{DBLP:conf/www/Wang0FNC18} and can be either structured \cite{DBLP:conf/recsys/Sun00BHX18} or semi-structured \cite{DBLP:conf/sigir/ZhangL0ZLM14,DBLP:conf/kdd/Bauman0T17,DBLP:conf/sigir/ChenQZX16}. Interestingly, in \cite{DBLP:journals/corr/abs-1804-11192} the authors argue about a new generation of knowledge-aware recommendation engines able to exploit information encoded in knowledge graphs (KG) to produce meaningful recommendations: ``\textit{For example, with knowledge graph about movies, actors, and directors, the system can explain to the user a movie is recommended because he has watched many movies starred by an actor}''.

In this work, we propose a \textit{knowledge-aware Hybrid Factorization Machine} (\framework) to train interpretable models in recommendation scenarios taking advantage of semantics-aware information (Section \ref{sec:approach}). \framework relies on Factorization Machines \cite{rendle2010factorization} and it extends them in different key aspects by making use of the semantic information encoded in a knowledge graph. 
With \framework we address the following research questions:
\vspace{-0.3cm}
\begin{description}
	\item[RQ1] Can we develop a model-based recommendation engine whose results are very accurate and, at the same time, interpretable with respect to an explicitly stated semantics coming from a knowledge graph?
	\item[RQ2] Can we evaluate that the original semantics of items features is preserved after the model has been trained?
	\item[RQ3] How to measure with an offline evaluation that the proposed model is really able to identify meaningful features by exploiting their explicit semantics?
\end{description}\vspace{-0.3cm}
We show how \framework may exploit data coming from knowledge graphs as side information to build a recommender system whose final results are accurate and, at the same time, semantically interpretable. 
With \framework, we build a model in which the meaning of each latent factor is bound to an explicit content-based feature extracted from a knowledge graph. Doing this, after the model has been trained, we still have an explicit reference to the original semantics of the features describing the items, thus making possible the interpretation of the final results. 
To answer RQ2, and RQ3 we introduce two metrics, Semantic Accuracy (SA@$K$) (Section \ref{approach_sa}) and Robustness (n-Rob@$K$) (Section \ref{approach_gr}), to measure the interpretability of a knowledge-aware recommendation engine.
The remainder of this paper is structured as follows: we evaluated \framework on two different publicly available datasets by getting content-based explicit features from data encoded in the \dbpedia knowledge graph. We analyzed the performance of the approach in terms of accuracy of results (Section \ref{sec:acc}) by exploiting categorical, ontological and factual features (see Section \ref{sec:approach}). 
Finally, we tested the robustness of \framework with respect to its interpretability (Sections \ref{sec:semacc}, and \ref{sec:semrob}) showing that it ranks meaningful features higher and is able to regenerate them in case they are removed from the original dataset.

\section{Knowledge-aware Hybrid Factorization Machines for Top-N Recommendation}\label{sec:background}
In this section, we briefly recap the main technologies we adopted to develop \framework. We introduce Vector Space Models for recommender systems, and then we give a quick overview of Factorization Machines (FM).

Content-based recommender systems rely on the assumption that it is possible to predict the future behavior of users based on their personalized profile. Profiles for users can be built by exploiting the characteristics of the items they liked in the past or some other available side information. Several approaches have been proposed, that take advantage of side information in different ways: some of them consider tags \cite{DBLP:conf/iui/VigSR09}, demographic data \cite{DBLP:journals/kais/ZhaoLHWWL16} or they extract information from collective knowledge bases \cite{DBLP:conf/recsys/NoiaMOR12} to mitigate the cold start problem \cite{DBLP:journals/umuai/Fernandez-Tobias19}.
Many of the most popular and adopted CB approaches make use of a Vector Space Model (VSM). In VSM users and items are represented by means of Boolean or weighted vectors. Their respective positions and the distance, or better the proximity, between them, provides a measure of how these two entities are related or similar. 
The choice of item features may substantially differ depending on their availability and application scenario: crowd-sourced tags, categorical, ontological, or textual knowledge are just some of the most exploited ones. 
All in all, in a CB approach we need (i) to get reliable items descriptions, (ii) a way to measure the strength of each feature for each item, (iii) to represent users and finally (iv) to measure similarities.
Regarding the first point,  nowadays we can easily get descriptions related to an item from the Web. In particular, thanks to the Linked Open Data initiative a lot of semantically structured knowledge is publicly available in the form of Linked Data datasets. 
%
%The space containing users and items vectors is composed by dimensions that may refer to characteristics of the users (e.g., the propensity to watch movies in the morning) or of the items as well as to their combination.
%
%\subsection{Knowledge Graphs and Linked Data}\label{sec:kg-and-lod}
%\input{src/approach_knowledgegraphs}

%\vspace{-0.5cm}
\subsection{From Factorization Machines to knowledge-aware Hybrid Factorization Machines}\label{sec:approach}
Factorization models have proven to be very effective in a recommendation scenario \cite{rendle2011context}. High prediction accuracy and the subtle modeling of user-item interactions let these models operate efficiently even in very sparse settings. 
Among all the different factorization models, factorization machines propose a unified general model to represent most of them.
Here we report the definition related to a factorization model of order 2 for a recommendation problem involving only implicit ratings. Nevertheless, the model can be easily extended to a more expressive representation by taking into account, e.g.,
  demographic and social information \cite{DBLP:reference/sp/AdomaviciusT15}, multi-criteria \cite{DBLP:reference/sp/AdomaviciusK15}, and even relations between contexts \cite{DBLP:conf/ijcai/ZhengMB15}.

\begin{figure}[ht]
	\centering
	\includegraphics[width=0.5\textwidth]{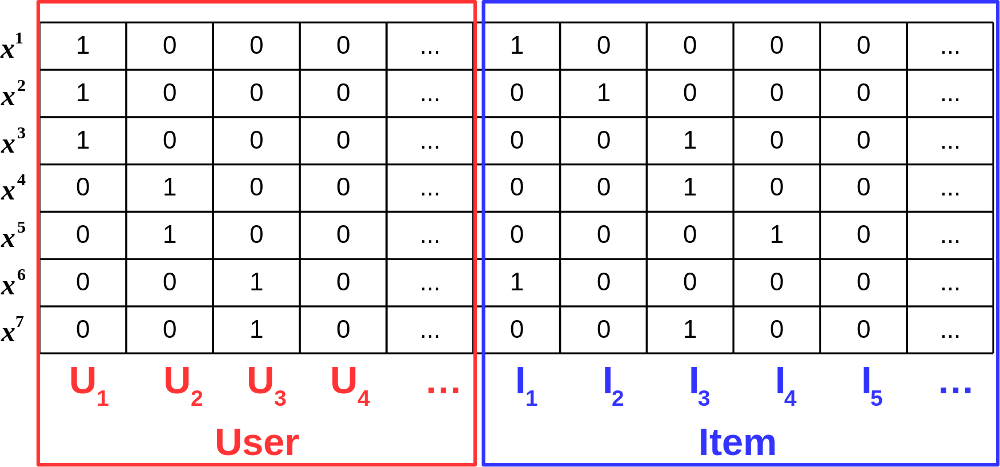}\vspace{-0.3cm}
	\caption{A visual representation of $\mathbf{X}$  for sparse real valued vectors $\mathbf{x^{ui}}$.}\vspace{-0.8cm}
	\label{fig:fm1}
		
\end{figure}

For each user $u \in U$ and each item $i \in I$ we build a binary vector $\mathbf{x^{ui}} \in \mathbb{R}^{1 \times n}$, with $n = |U|+|I|$, representing the interaction between $u$ and $i$ in the original user-item rating matrix. In this modeling, $\mathbf{x^{ui}}$ contains only two 1 values corresponding to $u$ and $i$ while all the other values are set to 0 (see Fig. \ref{fig:fm1}).
We then denote with $\mathbf{X} \in \mathbb{R}^{n \times m}$ the matrix containing as rows all possible $\mathbf{x^{ui}}$ we can build starting from the original user-item rating matrix as shown in Fig. \ref{fig:fm1}.
The FM score for each vector $\mathbf{x}$ is defined as:\vspace{-0.3cm}
%%%%%%%%%%%%%%%%%%%%%%%%%%%
\footnotesize
\begin{equation}\label{eq:fm1}\vspace{-0.3cm}
\hat{y}(\mathbf{x^{ui}}) = w_0 + \sum^{n}_{j=1}w_j\cdot x_j + \sum^{n}_{j=1}\sum^{n}_{p=j+1}x_j\cdot x_{p}\cdot\sum_{f = 1}^{k}v_{(j,f)}\cdot v_{(p,f)}
\end{equation}
\normalsize
%%%%%%%%%%%%%%%%%%%%%%%%%%%  
where the parameters to be learned are: $w_0$ representing the global bias; $w_j$ giving the importance to every single $x_j$; the pair $v_{(j,f)}$ and $v_{(p,f)}$ in $\sum_{f = 1}^{k}v_{(j,f)}\cdot v_{(p,f)}$ measuring the strength of the interaction between each pair of variables: $x_j$ and $x_{p}$. The number of latent factors is represented by $k$. This value is usually selected at design time when implementing the FM.

\begin{figure}[ht]
	\centering
	\includegraphics[width=0.5\textwidth]{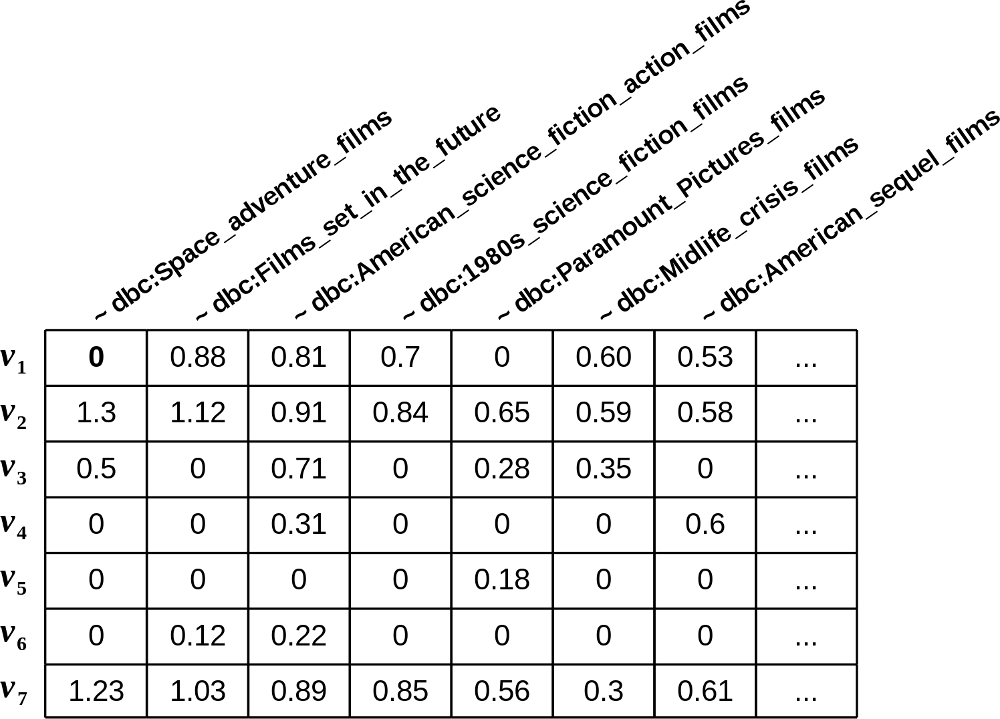}
	\caption{Example of real valued feature vectors for different items $v_j$. For lack of space we omitted the predicate \textit{dcterms:subject}}
	\label{fig:fm2}
\end{figure}

In order to make the recommendation results computed by \framework as semantically interpretable, we inject the knowledge encoded within a knowledge graph in a Factorization Machine. 
In a knowledge graph, each triple represents the  connection $\sigma \xrightarrow{\rho} \omega$ between two nodes, named \textit{subject} ($\sigma$) and  \textit{object} ($\omega$), through the \textit{relation} (\textit{predicate}) $\rho$. 
Given a set of features retrieved from a KG \cite{DBLP:conf/esws/NoiaMMPR18} we first bind them to the latent factors and then, since we address a Top-N recommendation problem, we train the model by using a Bayesian Personalized Ranking (BPR) criterion that takes into account entities within the original knowledge graph. 
In \cite{DBLP:conf/i-semantics/NoiaMORZ12}, the authors originally proposed to encode a Linked Data knowledge graph in a vector space model to develop a CB recommender system. 
Given a set of items $I = \{i_1,i_2, \hdots ,i_N \}$ in a catalog and their associated triples $\langle i, \rho, \omega\rangle$ in a knowledge graph $\mathcal{KG}$, we may build the set of all possible features as $F = \{\langle \rho, \omega \rangle \mid \langle i, \rho, \omega\rangle \in \mathcal{KG} \text{ with } i\in I \}$. Each item can be then represented as a vector of weights $\mathbf{i} = [v_{(i,1)},\ldots,v_{(i,\langle\rho,\omega\rangle)},\ldots, v_{(i,|F|)}]$, where $v_{(i,\langle\rho,\omega\rangle)}$ is computed as the normalized TF-IDF value for $\langle\rho,\omega\rangle$ as follows:\vspace{-0.1cm}
%\footnotesize 
%\begin{eqnarray}
%v_{(i,\langle\rho,\omega\rangle)} & = &  \underbrace{
%	\frac{|\{\langle \rho ,\omega\rangle \mid \langle i , \rho, \omega\rangle \in \mathcal{KG}\}|}
%	{\sqrt{\sum\limits_{\langle \rho,\omega\rangle \in F} |\{\langle \rho, \omega\rangle \mid \langle i , \rho, \omega \rangle \in \mathcal{KG}\}|^2}}}_{TF^{\mathcal{KG}}
%} \cdot \nonumber\\
%& \cdot &  \underbrace{
%	\log{\frac{|I|}{|\{j \mid \langle j,\rho,\omega\rangle \in \mathcal{KG} \text{ and } j \in I\}|}} }_{IDF^{\mathcal{KG}}}  \label{eq:tf-idf}
%\end{eqnarray}
%\normalsize
\footnotesize 
\begin{eqnarray}
v_{(i,\langle\rho,\omega\rangle)} = \underbrace{
	\frac{|\{\langle \rho ,\omega\rangle \mid \langle i , \rho, \omega\rangle \in \mathcal{KG}\}|}
	{\sqrt{\sum\limits_{\langle \rho,\omega\rangle \in F} |\{\langle \rho, \omega\rangle \mid \langle i , \rho, \omega \rangle \in \mathcal{KG}\}|^2}}}_{TF^{\mathcal{KG}}
} \cdot \underbrace{
	\log{\frac{|I|}{|\{j \mid \langle j,\rho,\omega\rangle \in \mathcal{KG} \text{ and } j \in I\}|}} }_{IDF^{\mathcal{KG}}}  \label{eq:tf-idf}
\end{eqnarray}
\normalsize
Since the numerator of $TF^{\mathcal{KG}}$ can only take values 0 or 1 and, each feature under the root in the denominator has value 0 or 1,  $v_{(i,\langle\rho,\omega\rangle)}$ is zero if $\langle\rho,\omega\rangle \not\in \mathcal{KG}$, and otherwise:\vspace{-0.2cm}
\footnotesize 
\begin{eqnarray}
v_{(i,\langle\rho,\omega\rangle)} = 
	\frac{\log{|I|}- \log{|\langle j,\rho,\omega\rangle \cap \mathcal{KG} | j \in I|} }
	{\sqrt{\sum\limits_{\langle \rho,\omega\rangle \in F} |\{\langle \rho, \omega\rangle \mid \langle i , \rho, \omega \rangle \in \mathcal{KG}\}|}}
\label{eq:tf-idf-2}
\end{eqnarray}
\normalsize
Analogously, when we have a set $U$ of users, we may represent them using the features describing the items they enjoyed in the past. In the following, when no confusion arises, we use $f$ to denote a feature $\langle\rho, \omega\rangle \in F$.
Given a user $u$, if we denote with $I^u$ the set of the items enjoyed by $u$, we may introduce the vector $\mathbf{u}=[v_{(u,1)},\ldots,v_{(u,f)}\ldots,v_{(u,|F|)}]$, where $v_{(u,f)}$ is:
\footnotesize
\[
v_{(u,f)} = \frac{
	\sum\limits_{i\in I^u} v_{(i,f)}
}{
	|\{i \mid i\in I^u \text{ and }v_{(i,f)} \not= 0\}|
} 
\]
\normalsize
Given the vectors $\mathbf{u}_j$, with $j \in [1\ldots|U|]$, and $\mathbf{i}_{p}$, with $p \in [1\ldots|I|]$, we build the matrix $\mathbf{V} \in \mathbb{R}^{n\times |F|}$ (see Fig. \ref{fig:fm2}) where the first $|U|$ rows have a one to one mapping with $\mathbf{u}_j$ while the last ones correspond to  $\mathbf{i}_{p}$. If we go back to Equation (\ref{eq:fm1}) we may see that, for each $\mathbf{x}$, the term $\sum^{n}_{j=1}\sum^{n}_{p=j+1}x_j\cdot x_{j'}\cdot\sum_{f = 1}^{k}v_{(j,f)}\cdot v_{(p,f)}$ is not zero only once, i.e., when both $x_j$ and $x_{p}$ are equal to 1. In the matrix depicted in Fig. \ref{fig:fm1}, this happens when there is an interaction between a user and an item.
Moreover, the summation $\sum_{f = 1}^{k}v_{(j,f)}\cdot v_{(p,f)}$ represents the dot product between two vectors: $\mathbf{v}_j$ and $\mathbf{v}_{p}$ with a size equal to $k$. Hence, $\mathbf{v}_j$ represents a latent representation of a user, $\mathbf{v}_{p}$ that of an item within the same latent space, and their interaction is evaluated through their dot product.

In order to inject the knowledge coming from $\mathcal{KG}$ into \framework, we keep Equation (\ref{fig:fm1}) and we set $k = |F|$. In other words, we impose a number of latent factors equal to the number of features describing all the items in our catalog. We want to stress here that our aim is not representing each feature through a latent vector, but to associate each factor to an explicit feature, obtaining latent vectors that are composed by explicit semantic features. Hence, we initialize the parameters  $\mathbf{v}_{j}$ and  $\mathbf{v}_{p}$ with their corresponding rows from $\mathbf{V}$ which in turn represent respectively $\mathbf{u}_j$ and $\mathbf{i}_{p}$. In this way, we try to identify each latent factor with a corresponding explicit feature. The intuition is that after the training phase, the resulting matrix $\hat{\mathbf{V}}$ still refers to the original features but contains better values for $v_{(j,f)}$ and $v_{(p,f)}$ that take into account also the latent interactions between users, items and features. It is noteworthy that after the training phase $\mathbf{u}_j$ and $\mathbf{i}_{p}$ (corresponding to $v_{(j,f)}$ and $v_{(p,f)}$ in $\mathbf{V}$) contain non-zero values also for features that are not originally in the description of the user $u$ or of the item $i$.
We extract the items vectors $\mathbf{v}_j$ from the matrix $\mathbf{\hat{V}}$, with the associated optimal values and we use them to implement an Item-kNN recommendation approach.  We measure similarities between each pair of items $i$ and $j$ by evaluating the cosine similarity of their corresponding vectors in $\mathbf{\hat{V}}$:
\footnotesize
\[\label{eqn:cvs}
cs(i,j) = \frac{
	\mathbf{v}_i \cdot \mathbf{v}_j
}
{
	\parallel \mathbf{v}_i \parallel \cdot \parallel \mathbf{v}_j \parallel
}
\]
\normalsize
Let us define $N^i$ as the set of neighbors for the item $i$, composed by the items which are more similar to $i$ according to the selected similarity measure. denoted as $N^i$. It is possible to choose $i$ such that $i \not\in I^u$ and a user $u$, we predict the score assigned by $u$ to $i$ as\vspace{-0.1cm}
\footnotesize
\begin{equation}\label{eqn:score}
score(u,i) = \frac{\sum\limits_{j \in N^i \cap I^u}{cs(i,j)}}{\sum\limits_{j \in N^i}{cs(i,j)}}\vspace{-0.1cm}
\end{equation}
\normalsize
Factorization machines can be easily trained to reduce the prediction error via gradient descent methods, alternating least-squares (ALS) and MCMC. Since we formulated our problem as a \TopN recommendation task, \framework can be trained using a learning to rank approach like Bayesian Personalized Ranking Criterion (BPR) \cite{DBLP:conf/uai/RendleFGS09}. 
The BPR criterion is optimized using a stochastic gradient descent algorithm on a set $D_S$ of triples $(u,i,j)$, with $i \in I^u$ and $j \not\in I^u$, selected through a random sampling from a uniform distribution.
Once the training phase returns the optimal model parameters, the item recommendation step can take place. 

In an RDF knowledge graph, we usually find different types of encoded information. 
\begin{itemize}
	\item \textbf{Factual.} This refers to statements such has \textit{The Matrix was directed by the Wachowskis} or \textit{Melbourne is located in Australia} when we describe attributes of an entity;
	\item \textbf{Categorical.} It is mainly used to state something about the subject of an entity. In this direction, the categories of Wikipedia pages are an excellent example. Categories can be used to cluster entities and are often organized hierarchically thus making possible to define them in a more generic or specific way;
	\item \textbf{Ontological.} This is a more restrictive and formal way to classify entities via a hierarchical structure of classes. Differently from categories, sub-classes and super-classes are connected through IS-A (transitive) relations. 
\end{itemize}
In Table \ref{tbl:khan} we show an example for categorical values obtained after the training (in the column \framework) together with the original TF-IDF ones computed for a movie from the Yahoo! Movies\footnote{\url{http://research.yahoo.com/Academic_Relations}} dataset.
\begin{table}[htbp]
	\centering
	\scriptsize
	\begin{tabular}{|r|r|ll|}
		\hline
		\multicolumn{1}{|l|}{\textbf{\framework}} & {\textbf{TF-IDF}} & \textbf{Predicate} & \textbf{Object} \\ \hline
		1.3669 & 0.2584 & dct:subject & dbc:Space\_adventure\_films \\ 
		1.1252 & 0.2730 & 	dct:subject & dbc:Films\_set\_in\_the\_future \\ 
		0.9133 & 0.2355 & 	dct:subject & dbc:American\_science\_fiction\_action\_films \\ 
		0.8485 & 0.3190 & 	dct:subject & dbc:1980s\_science\_fiction\_films \\ 
		0.6529 & 0.1549 & 	dct:subject & dbc:Paramount\_Pictures\_films \\ 
		0.5989 & 0.3468 & 	dct:subject & dbc:Midlife\_crisis\_films \\ 
		0.5940 & 0.1797 & 	dct:subject & dbc:American\_sequel\_films \\ 
		0.5862 & 0.2661 & 	dct:subject & dbc:Film\_scores\_by\_James\_Horner \\ 
		0.5634 & 0.2502 & 	dct:subject & dbc:Films\_shot\_in\_San\_Francisco \\ 
		0.5583 & 0.1999 & 	dct:subject & dbc:1980s\_action\_thriller\_films \\ 
		\hline
	\end{tabular}
	\caption{Top-10 features computed by \framework for the movie \texttt{"Star Trek II - The Wrath of Khan"}.}
	\label{tbl:khan}
\end{table}
%\vspace{-0.2cm}
\section{Semantic Accuracy and Generative Robustness}\label{sec:semantics}
The proposed approach let us keep the meaning of the ``latent'' factors computed via a factorization machine thus making possible an interpretation of the recommended results. 
To assess that \framework preserves the semantics of the features in $\mathbf{V}$ after the training phase, we propose an automated offline procedure to measure \textit{Semantic Accuracy}.
Moreover, we define as \textit{Robustness} the ability to assign a higher value to important features after one or more feature removals.
\subsection{Semantic Accuracy}\label{approach_sa}
The main idea behind Semantic Accuracy is to evaluate, given an item $i$, how well \framework is able to return its original features available in the computed top-K list $\mathbf{v}_i$. In other words, given the set of features of $i$ represented by $F^i = \{f_{1}^i, \ldots,f_{m}^i, \ldots f_{M}^i\}$, with $F^i \subseteq F$, we check if the values in $\mathbf{v}_i$, corresponding to $f_{m,i} \in F^i$, are higher than those corresponding to $f \not\in F^i$.
For the set of $M$ features initially describing $i$  we see how many of them appear in the set $top(\mathbf{v}_i,M)$ representing  the top-$M$ features in $\mathbf{v}_i$. We then normalize this number by the size of $F^i$ and average on all the items within the catalog $I$.
\footnotesize
\[
\texttt{Semantic Accuracy (SA@$M$)} = \frac{\sum\limits_{i \in I} {\frac{|top(\mathbf{v}_i,M)\cap F^i|}{|F^i|}}}{|I|}
\]
\normalsize
In many practical scenarios we may have $|F| \gg M$. Hence, we might also be interested in measuring the accuracy for different sizes of the top list.
Since items could be described with a  different number of features, the size of the top list could be a function of the original size of the item description.
Thus, we measured \texttt{SA@$nM$} with $n \in \{1,2,3,4,5, \ldots\}$ and evaluate the number of  features in $F^i$ available in the top-$n\cdot M$ elements of $\mathbf{v}_i$.\vspace{-0.2cm}
\footnotesize
\[
\texttt{SA@$nM$} = \frac{\sum\limits_{i \in I} {\frac{|top(\mathbf{v}_i,n \cdot M)\cap F^i|}{|F^i|}}}{|I|}
\]
\normalsize
\vspace{-1cm}
\subsection{Robustness}\label{approach_gr}
\vspace{-0.1cm}
Although \texttt{SA@nM} may result very useful to understand if \framework assigns weights according to the original description of item $i$, we still do not know if a high value in $\mathbf{v}_i$ really means that the corresponding feature is important to define $i$. In other words, are we sure that \framework promotes important features for $i$?

\noindent In order to provide a way to measure such ``meaningfulness'' for a given feature, we suppose, for a moment, that a particular feature ${\langle \rho , \omega \rangle}$ is useful to describe an item $i$ but the corresponding triple $\langle i, \rho , \omega \rangle$ is not represented in the knowledge graph. In case \framework was robust in generating weights for unknown features, it should discover the importance of that feature and modify its value to make it enter the Top-$K$ features in ${\mathbf{v}}_i$. 
Starting from this observation, the idea to measure robustness is then to ``forget'' a triple involving $i$ and check if \framework can generate it. 
In order to implement such process we proceed by following these steps:\vspace{-0.3cm}
\begin{itemize}
	\item we  train \framework thus obtaining optimal values $v_i$ for all the features in $F^i$;
	\item the feature $f^i_{MAX} \in F^i$ with the highest value in $v_i$ is identified;
	\item we retrain the model again initializing $f^i_{MAX} = 0$ and we compute $v'_i$.
\end{itemize}\vspace{-0.2cm}
After the above steps, if $f^i_{MAX} \in top(v'_i,M)$ then we can say that \framework shows a high robustness in identifying important features. 
Given a catalog $I$, we may then define the \textit{Robustness for 1 removed feature @M} \texttt{(1-Rob@M)} as the number of items for which $f^i_{MAX} \in top(v'_i,M)$ divided by the size of $I$.\vspace{-0.2cm}
\footnotesize
\[
\texttt{1-Rob@M} = \frac{\sum\limits_{i \in I}{|\{ i \mid f^i_{MAX} \in top(v'_i,M)\}|}}{|I|}
\]
\normalsize
\noindent Similarly to \texttt{SA@$nM$}, we may define \texttt{1-Rob@nM}.
%\vspace{-0.4cm}
\section{Experimental Evaluation}\label{sec:eval}
In this section, we will detail three distinct experiments. We specifically designed them to answer the research questions posed in Section 1. In details, we want to assess if: i) \framework's recommendations are accurate; ii) \framework generally preserves the semantics of original features; iii) \framework promotes significant features.
\noindent\textbf{Datasets. } To provide an answer to our research questions, we evaluated the performance of our method on two well-known datasets for recommender systems belonging to movies domain. 
\yahoo (Yahoo! Webscope dataset ydata-ymovies-user-movie-ratings-content-v1\_0)\footnote{\url{http://research.yahoo.com/Academic_Relations}} contains movies ratings generated on Yahoo! Movies up to November 2003. It provides content, demographic and ratings information on a [1..5] scale, and mappings to \movielens and \texttt{EachMovie} datasets. \fbmovies dataset has been released for the Linked Open Data challenge co-located with ESWC 2015\footnote{\url{https://2015.eswc-conferences.org/program/semwebeval.html}}. Only implicit feedback is available for this dataset, but for each item a link to DBpedia is provided. 
To map items in \yahoo and other well-known datasets, we extracted all the updated items-features mappings and we made them publicly available\footnote{\url{https://github.com/sisinflab/LinkedDatasets/}}. 
%https://anonymous.4open.science/repository/bb5320de-d2f0-4816-9507-0079354c97e6/
%https://github.com/sisinflab/LinkedDatasets/
%https://anonymous.4open.science/repository/84f142bc-85ca-4c2a-9df3-5a69b26393b1/
Datasets statistics are shown in Table \ref{tbl:datastats}.
%\setlength\tabcolsep{3.0pt} 
%\begin{table}[hbt!]
%	%	\centering
%	\scriptsize
%	\begin{tabular}{l r r r r r}
%		\hline
%		Dataset & \#Users &  \#Items &  \#Transactions & \#Features  & Sparsity\\
%		\hline
%		Yahoo! Movies & 4000 &  2,626 & 69,846 & 988,734 &  99.34\%\\ 
%		LibraryThing & 7223 &  11,695 & 410,210 & 183,182 &  99.51\%\\
%		Last FM & 1375 &  8,289 & 60,701 & 434,817 & 99.47\%\\
%		Facebook Music & 52068 & 5,749 & 1,374,994 & 345,452 & 99.54\%\\
%		Facebook Movies & 32143 & 3,901 & 689,561 & 180,573 & 99.45\%\\
%		Facebook Books & 1398 & 2,933 & 18,978 & 111,401 & 99.53\%\\
%		\hline
%	\end{tabular}
%	\caption{Datasets statistics.}
%%	\vspace{-1cm}
%	\label{tbl:datastats}
%\end{table}
%\vspace{-0.5cm}
\setlength\tabcolsep{3.0pt} 
\begin{table}[hbt!]
		\centering
%	\scriptsize
	\begin{tabular}{l r r r r r}
		\hline
		Dataset & \#Users &  \#Items &  \#Transactions & \#Features  & Sparsity\\
		\hline
		Yahoo! Movies & 4000 &  2,626 & 69,846 & 988,734 &  99.34\%\\ 
		Facebook Movies & 32143 & 3,901 & 689,561 & 180,573 & 99.45\%\\
		\hline
	\end{tabular}
	\caption{Datasets statistics.}\vspace{-0.3cm}
	\label{tbl:datastats}
\end{table}
\\\noindent\textbf{Experimental Setting. } 
"All Unrated Items" \cite{steck2013evaluation} protocol has been adopted to compare different algorithms. 
%In All Unrated Items, for each user, all the items that have not yet been rated by the user all over the catalog are considered. 
We have split the dataset using Hold-Out 80-20 retaining for every user the 80\% of their ratings in the training set and the remaining 20\% in the test set. Moreover, a temporal split has been performed \cite{DBLP:reference/sp/GunawardanaS15} whenever timestamps associated to every transaction is available. 
%https://anonymous.4open.science/repository/29ee0236-d48c-407e-86a7-d8f4b718ff80/
%https://github.com/sisinflab/HybridFactorizationMachines/

%\noindent\textbf{Features extraction. }The feature extraction is one of the most sensitive steps in our approach. A wrong feature selection may result in noisy data, or in the lack of some important features. This preprocessing was basically divided in three steps: (i) "\textbf{Extraction}", in which we retrieve data from the DBpedia knowledge graph, (ii) "\textbf{Selection}" where only features involved in the specific experiment are selected, and (iii) "\textbf{Filtering}" in which uninformative features are removed \cite{DBLP:conf/esws/NoiaMMPR18}. 
\noindent\textbf{Extraction}. Thanks to the publicly available mappings, all the items from the datasets represented in Table \ref{tbl:datastats} come with a \dbpedia link. Exploiting this reference, we retrieved all the $\langle \rho, \omega \rangle$ pairs. Some noisy features (based on the following predicates) have been excluded: \texttt{owl:sameAs}, \texttt{dbo:thumbnail}, \texttt{foaf:depiction}, \texttt{prov:wasDerivedFrom}, \texttt{foaf:isPrimaryTopicOf}.
\noindent\textbf{Selection}. We performed our experiments with three different settings to analyze the impact of the different kind of features. The features have been chosen as they are present in all the different domains and because of their factual, categorical or ontological meaning:
\vspace{-0.2cm}
\begin{itemize}
	\item \textbf{Categorical Setting (CS)}: We selected only the features containing the property \texttt{dcterms:subject}. 
	\item \textbf{Ontological Setting (OS)}: In this case the only feature we considered is \texttt{rdf:type}.
	\item \textbf{Factual Setting (FS)}: We considered all the features but those involving the properties selected in OS, and CS. \\ 
\end{itemize}
\vspace{-0.5cm}
\noindent\textbf{Filtering}. This last step corresponds to the removal of irrelevant features, that bring little value to the recommendation task, but, at the same time, pose scalability issues. The pre-processing phase has been done following \cite{DBLP:conf/esws/NoiaMMPR18}, and \cite{DBLP:conf/wims/PaulheimF12} with a unique threshold. Thresholds (corresponding to \textit{tm} \cite{DBLP:conf/esws/NoiaMMPR18}, and \textit{p} \cite{DBLP:conf/wims/PaulheimF12} for missing values) and the considered features for each dataset are represented in Table \ref{tbl:features}.
\vspace{-0.7cm}
\setlength\tabcolsep{2.0pt} 
\begin{table}[htbp]
	\scriptsize
	\begin{tabular}{|l|r|r|r|r|r|r|r|}
		\hline
		\multicolumn{1}{|c|}{\textbf{}} & \multicolumn{1}{|c|}{\textbf{}} & \multicolumn{2}{|c|}{\textbf{Categorical Setting}}& \multicolumn{2}{|c|}{\textbf{Ontological Setting}} & \multicolumn{2}{|c|}{\textbf{Factual Setting}} \\ 
		\hline
		Datasets & \multicolumn{1}{l|}{Threshold} & \multicolumn{1}{l|}{Total} & \multicolumn{1}{r|}{Selected} & \multicolumn{1}{l|}{Total} & \multicolumn{1}{r|}{Selected} & \multicolumn{1}{l|}{Total} & \multicolumn{1}{r|}{Selected} \\ \hline
		\textbf{\yahoo} & 99.62 & 26155 & 747 & 38699 & 1240 & 950035 & 3186 \\ \hline
%		\textbf{\library} & 99.91 & 9443 & 1169 & 14585 & 1934 & 168597 & 5826 \\ \hline
%		\textbf{\lastfm} & 99.88 & 16422 & 1315 & 30734 & 3032 & 404083 & 9413 \\ \hline
%		\textbf{\fbmusic} & 99.83 & 15016 & 1057 & 27988 & 2531 & 317464 & 7881 \\ \hline
		\textbf{\fbmovies} & 99.74 & 8843 & 1103 & 13828 & 1848 & 166745 & 5427 \\ \hline
%		\textbf{\fbbooks} & 99.66 & 6231 & 263 & 9881 & 592 & 101520 & 1315 \\ \hline
	\end{tabular}
	\caption{Considered features in the different settings}
	\vspace{-1.2cm}
	\label{tbl:features}
\end{table}\vspace{-0.6cm}

\subsection{Accuracy Evaluation}\label{sec:acc}
The goal of this evaluation is to assess if the controlled injection of \texttt{Linked Data} positively affects the training of Factorization Machines. For this reason, \framework is not compared with other state-of-art interpretable models but with only the algorithms that are more related to our approach. We compared \framework\footnote{\url{https://github.com/sisinflab/HybridFactorizationMachines}}  w.r.t. a canonical 2 degree Factorization Machine (users and items are intended as features of the original formulation) by optimizing the recommendation list ranking via BPR (BPR-FM). In order to preserve the expressiveness of the model, we used the same number of hidden factors (see the "Selected" column in Table \ref{tbl:features}). Since we use items similarity in the last step of our approach (see Equation (\ref{eqn:score})), we compared \framework against an \textit{Attribute Based Item-kNN} (ABItem-kNN) algorithm, where each item is represented as a vector of weights, computed through a TF-IDF model. In this model, the attributes are computed via Equation (\ref{eq:tf-idf}).
We also compared \framework also against a pure Item-kNN, that is an item-based implementation of the k-nearest neighbors algorithm. It finds the k-nearest item neighbors based on Pearson Correlation. 
https://github.com/sisinflab/HybridFactorizationMachines
%https://anonymous.4open.science/repository/9c81680f-e268-434e-8043-48cb4eb1f9c5/
Regarding BPR parameters, \textit{learning rate}, \textit{bias regularization}, \textit{user regularization}, \textit{positive item regularization}, and \textit{negative item regularization} have been set respectively to 0.05, 0, 0.0025, 0.0025 and 0.00025 while a sampler "without replacement" has been adopted in order to sample the triples as suggested by authors \cite{DBLP:conf/uai/RendleFGS09}.
We compared \framework also against the corresponding User-based nearest neighbor scheme, and Most-Popular, a simple baseline that shows high performance in specific scenarios \cite{DBLP:conf/recsys/CremonesiKT10}. In our context, we considered mandatory to also compare against a pure knowledge-graph content-based baseline based on Vector Space Model ($VSM$) \cite{DBLP:conf/i-semantics/NoiaMORZ12}.
In order to evaluate our approach, we measured accuracy through Precision@N ($Prec@N$) and Normalized Discounted Cumulative Gain ($nDCG@N$). 
The evaluation has been performed considering Top-10 
%\cite{DBLP:conf/icml/CaoQLTL07}
\cite{DBLP:conf/recsys/CremonesiKT10}
%\cite{DBLP:conf/recsys/ShiLH10}
 recommendations for all the datasets. When a rating score was available (\yahoo), a \textit{Threshold-based relevant items} condition \cite{DBLP:journals/umuai/CamposDC14,DBLP:conf/um/AnelliBNBTS17} was adopted with a relevance threshold of 4 over 5 stars in order to take into account only relevant items. 
\begin{figure}[h]
%	\centering
	\begin{minipage}{.57\textwidth}\vspace{+0.3cm}
		\scriptsize
		\begin{tabular}{|l|r|r|r|}
			\hline
			\multicolumn{1}{|c|}{\textbf{}} & \multicolumn{1}{c|}{\texttt{Facebook}} & \multicolumn{2}{c|}{\texttt{Yahoo!}} \\ 
			\hline
			Categorical Setting (CS)& \multicolumn{1}{l|}{P@10} & \multicolumn{1}{l|}{P@10}  & \multicolumn{1}{l|}{nDCG@10}  \\ \hline
			\textbf{ABItem-kNN} & 0.0173$^*$ & \underline{0.0421}$^*$ & \underline{0.1174}$^*$ \\ \hline
			\textbf{BPR-FM} & 0.0158$^*$ & 0.0189$^*$ & 0.0344$^*$  \\ \hline
			\textbf{MostPopular} & 0.0118$^*$ & 0.0154$^*$ & 0.0271$^*$  \\ \hline
			\textbf{ItemKnn} & \underline{0.0262}$^*$ & 0.0203$^*$ & 0.0427$^*$  \\ \hline
			\textbf{UserKnn} & 0.0168$^*$ & 0.0231$^*$ & 0.0474$^*$  \\ \hline
			\textbf{VSM} & 0.0185$^*$ & 0.0385$^*$ & 0.1129$^*$  \\ \hline
			\textbf{kaHFM}  & \textbf{0.0296} & \textbf{0.0524} & \textbf{0.1399}  \\ \hline \hline
			Ontological Setting (OS)& \multicolumn{1}{l|}{P@10} & \multicolumn{1}{l|}{P@10}  & \multicolumn{1}{l|}{nDCG@10}  \\ \hline
			\textbf{ABItem-kNN} & 0.0172 & \underline{0.0427}$^*$ & \underline{0.1223}$^*$  \\ \hline
			\textbf{BPR-FM} & 0.0155$^*$ & 0.0199$^*$ & 0.0356$^*$  \\ \hline
			\textbf{MostPopular} & 0.0118$^*$ & 0.0154$^*$ & 0.0271$^*$  \\ \hline
			\textbf{ItemKnn} & \underline{0.0263}$^*$ & 0.0203$^*$ & 0.0427$^*$  \\ \hline
			\textbf{UserKnn} & 0.0168$^*$ & 0.0232$^*$ & 0.0474$^*$  \\ \hline
			\textbf{VSM} & 0.0181$^*$ & 0.0349$^*$ & 0.1083$^*$ \\ \hline
			\textbf{kaHFM} & \textbf{0.0273} & \textbf{0.0521} & \textbf{0.1380}  \\ \hline \hline
			Factual Setting (FS)& \multicolumn{1}{l|}{P@10} & \multicolumn{1}{l|}{P@10}  & \multicolumn{1}{l|}{nDCG@10}  \\ \hline
			\textbf{ABItem-kNN} & 0.0234 & \underline{0.0619} & \textbf{0.1764}  \\ \hline
			\textbf{BPR-FM} & 0.0157 & 0.0177 & 0.0305 \\ \hline
			\textbf{MostPopular} & 0.0123  & 0.0154 & 0.0271  \\ \hline
			\textbf{ItemKnn} & \textbf{0.0273} & 0.0203 & 0.0427  \\ \hline
			\textbf{UserKnn} & 0.0176 & 0.0232 & 0.0474 \\ \hline
			\textbf{VSM} & 0.0219 & \textbf{0.0627} & \underline{0.1725}  \\ \hline
			\textbf{kaHFM} & \underline{0.0240} & 0.0564 & 0.1434  \\ \hline
		\end{tabular}\vspace{+0.8cm}
%		\caption{Accuracy results for \fbmovies, and \yahoo}\label{fig:acc_res}\vspace{-0.6cm}
		
%		\captionof{table}[text for list of tables]{A table right of a figure}
	\end{minipage}
	\begin{minipage}{.44\textwidth}%\vspace{+0.1cm}
%	\begin{figure}
		\begin{flushright}
			\begin{tabular}{c}
				%		\subfloat[Precision on \library]{\includegraphics[width=4.2cm]{imgs/imgs_withNewBaselines/library.png}}\vspace{-0.1cm} & 
				\subfloat[\yahoo]{\includegraphics[width=5cm]{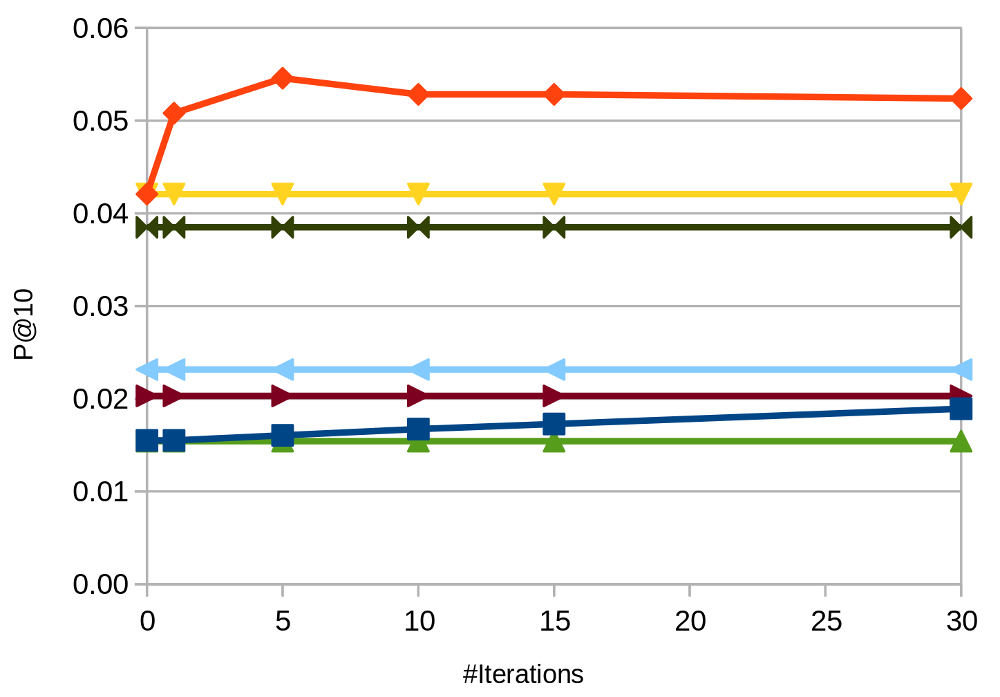}}\vspace{-0.3cm}   \\
				%		\subfloat[Precision on \lastfm]{\includegraphics[width=4.2cm]{imgs/imgs_withNewBaselines/lastfm.png}}\vspace{-0.1cm} &
				%		\multirow{-3}{*}{\subfloat[legend]{\includegraphics[width=2cm]{imgs/legend.png}}}\\
				\subfloat[\fbmovies]{\includegraphics[width=5cm]{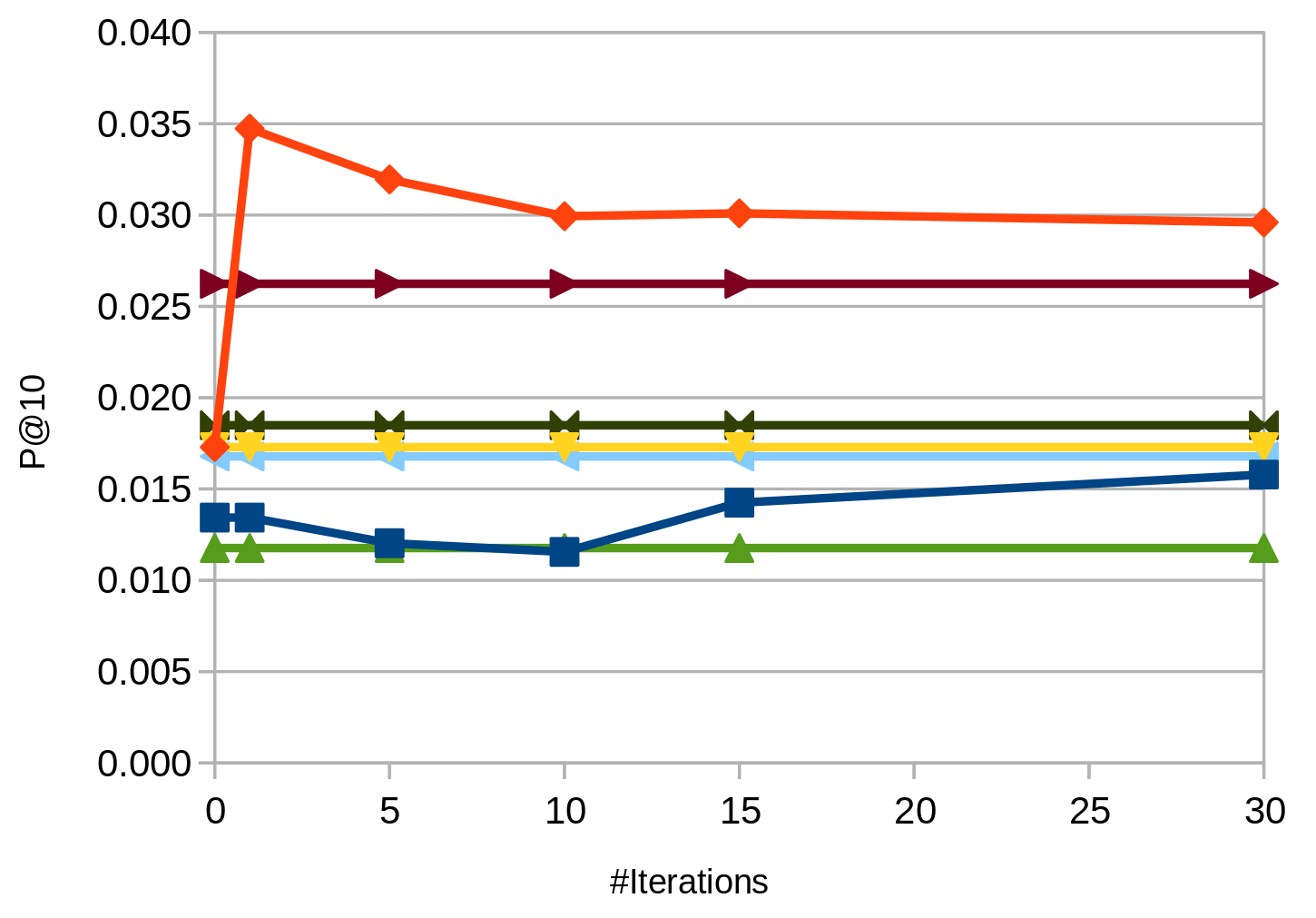}}\vspace{-0.1cm}  \\
				%		\subfloat[Precision on \fbmusic]{\includegraphics[width=4.2cm]{imgs/imgs_withNewBaselines/facebookmusic.png}}\vspace{-0.1cm} & 
				%		\subfloat[Precision on \fbbooks]{\includegraphics[width=4.2cm]{imgs/imgs_withNewBaselines/facebookbook.png}}\vspace{-0.1cm} \\ 
			\end{tabular}
			\begin{tabular}{ccc}
				%		\subfloat[legend]{\includegraphics[width=\linewidth]{imgs/imgs_withNewBaselines/hlegend.png}}\vspace{-0.2cm}\\
				\stackunder[5pt]{\includegraphics[height=8pt]{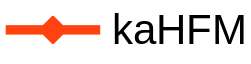}}{}
				\stackunder[5pt]{\includegraphics[height=8pt]{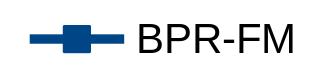}}{}
				\stackunder[5pt]{\includegraphics[height=8pt]{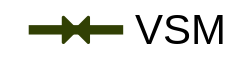}}{}
				\stackunder[5pt]{\includegraphics[height=8pt]{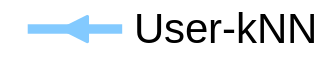}}{}\\
				\stackunder[5pt]{\includegraphics[height=8pt]{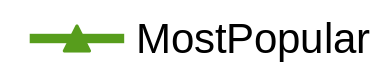}}{}
				\stackunder[5pt]{\includegraphics[height=8pt]{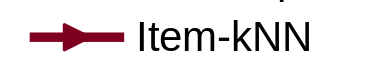}}{}
				\stackunder[5pt]{\includegraphics[height=8pt]{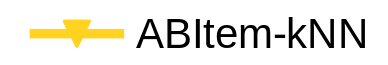}}{}\\
			\end{tabular}\vspace{-0.3cm}
%			\caption{Precision@10 varying \# iterations {0, 1, 5 , 10 , 15, 30} }\vspace{-0.3cm}
%			\label{fig:accuracy_results}
		\end{flushright}
%		\captionof{figure}[text for list of tables]{A table right of a figure}
%	\end{figure}
	\end{minipage}\hfill
	\caption{Accuracy results for \fbmovies, and \yahoo. In figures: Precision@10 varying \# iterations {0, 1, 5 , 10 , 15, 30} }\label{fig:acc_res}
\end{figure}
Fig. \ref{fig:acc_res} shows the results of our experiments regarding accuracy. 
In all the tables we highlight in \textbf{bold} the best result while we \underline{underline} the second one. Statistically significant results are denoted with a $^*$ mark considering  Student's paired t-test with a 0.05 level.
\yahoo experiments show that in Categorical and Ontological settings our method is the most accurate. In the \yahoo mapping, a strong popularity bias is present and it is interesting to notice that this affects only the Factual setting leading our approach to be less precise than ABItem-kNN.
In \texttt{Facebook Movies} we see very a good improvement in terms of accuracy as it almost doubles up the ABItem-kNN algorithm values. 
We compared \framework against ABItem-kNN to check if the collaborative trained features may lead to better similarity values. This hypothesis seems to be confirmed since in former experiments \framework beats ABItem-kNN in almost all settings. This suggests that collaborative trained features achieve better accuracy results. Moreover, we want to check if a knowledge-graph-based initialization of latent factors may improve the performance of Factorization Machines. \framework always beats BPR-FM, and in our opinion, this happens since the random initialization takes a while to drive the Factorization machine to reach good performance. Finally, we want to check if collaborative trained features lead to better accuracy results than a purely informativeness-based Vector Space Model even though it is in its knowledge-graph-aware version. This seems to be confirmed in our experiments, since \framework beats $VSM$ in almost all cases.
In order to strengthen the results we got, we computed recommendations with 0,1,5,10,15,30 iterations. For the sake of brevity we report here\footnote{Results of the full experiments: \url{https://github.com/sisinflab/papers-results/tree/master/kahfm-results/}} only the plots related to Categorical setting (Fig. \ref{fig:acc_res})
%https://anonymous.4open.science/repository/8bd244b6-91c4-443b-824c-cdccf66df4fe/
%https://anonymous.4open.science/repository/8bd244b6-91c4-443b-824c-cdccf66df4fe/
%https://github.com/sisinflab/papers-results/tree/master/kahfm-results/
It is worth to notice that in every case we considered, we show the best performance in one of these iterations. Moreover, the positive influence of the initialization of the feature vectors is particularly evident in all the datasets, with performances being very similar to the ones depicted in \cite{DBLP:conf/uai/RendleFGS09}. Given the obtained results we may say that the answer to RQ1 is positive when adopting \framework.
\subsection{Semantic Accuracy}\label{sec:semacc}
\vspace{-0.2cm}
The previous experiments showed the effectiveness of our approach in terms of accuracy of recommendation. In practical terms, we proved that: (i) content initialization generally lead to better performance with our method, (ii) the obtained items vectors are fine-tuned better than the original ones for a \TopN item recommendation task, (iii) results may depend on the features we extract from the Knowledge Graph. However, we still do not know if the original semantics of the features is preserved in the new space computed after the training of \framework (as we want to assess by posing RQ2). 
In Section \ref{approach_sa} we introduced \texttt{Semantics Accuracy ($SA@nM$)} as a metric to automatically check if the importance computed by \framework and associated to each feature reflects the actual meaning of that feature.
Thus, we measured \texttt{SA@$nM$} with $n \in \{1,2,3,4,5\}$ and $M = 10$, and evaluated the number of ground features available in the top-$nM$ elements of $\mathbf{v}_i$ for each dataset.
\begin{table}[ht]
	\centering
%	\scriptsize
	\begin{tabular}{|l|r|r|r|r|r| |r|}
		\hline
		
		Semantics Accuracy & \multicolumn{1}{l|}{SA@M} & \multicolumn{1}{l|}{SA@2M} & \multicolumn{1}{l|}{SA@3M} & \multicolumn{1}{l|}{SA@4M} & \multicolumn{1}{l||}{SA@5M} & \multicolumn{1}{l|}{F.A.} \\ \hline
		\textbf{\yahoo} & 0.847 & 0.863 & 0.865 & 0.868 & 0.873 & 12.143\\ \hline
%		\textbf{\library} & 0.960 & 0.996 & 0.998 & 0.999 & 0.999 & 3.820\\ \hline
%		\textbf{\lastfm} & 0.960 & 0.987 & 0.991 & 0.994 & 0.995 & 6.615\\ \hline
%		\textbf{\fbmusic} & 0.892 & 0.948 & 0.962 & 0.970 & 0.974 & 7.113\\ \hline
		\textbf{\fbmovies} & 0.864 & 0.883 & 0.889 & 0.894 & 0.899& 12.856\\ \hline
%		\textbf{\fbbooks} & 0.995 & 1 & 1 & 1 & 1 & 3.133\\ \hline
	\end{tabular}
	\caption{Semantics Accuracy results for different values of M. F.A. denotes the Feature Average number per item.}
	\label{tbl:meaningfulness}
\end{table}
Table \ref{tbl:meaningfulness} shows the results for all the different datasets computed in the Categorical setting. In general, the results we obtain are noteworthy. We now examine the worst one to better understand the actual meaning of the values we get. In \yahoo, 747 different features compose each item vector (see Table \ref{tbl:features}). After the training phase, on average, more than 10 (equal to $0.847 \times 12.143$) over 12 features (last column in Table \ref{tbl:meaningfulness}) are equal to the original features list. This means that \framework was able to compute almost the same features starting from hundreds of them. Also in this case, given the obtained results we may provide a positive answer to RQ2.
\subsection{Generative Robustness}\label{sec:semrob}
The previous experiment showed that the features computed by \framework keep their original semantics if already present in the item description. 
In section \ref{approach_gr}, we introduced a procedure to measure the capability of \framework to compute meaningful features. Here, we computed \texttt{1-Rob@nM} for the two adopted datasets. Results are represented in Table \ref{tbl:robustness}.
\setlength\tabcolsep{0.1pt}
\begin{table}[ht]
	\centering
%	\scriptsize
	\begin{tabular}{|l|r|r|r|r|r| |r|}
		\hline	
		1-Robustness & \multicolumn{1}{l|}{\texttt{1-Rob}@M} & \multicolumn{1}{l|}{\texttt{1-Rob}@2M} & \multicolumn{1}{l|}{\texttt{1-Rob}@3M} & \multicolumn{1}{l|}{\texttt{1-Rob}@4M} & \multicolumn{1}{l||}{\texttt{1-Rob}@5M} & \multicolumn{1}{l|}{F.A.} \\ \hline
		\textbf{\yahoo} & 0.487 & 0.645 & 0.713 & 0.756 & 0.793 & 12.143\\ \hline
%		\textbf{\library} & 0.275 & 0.481 & 0.554 & 0.597 & 0.632 & 3.820\\ \hline
%		\textbf{\lastfm} & 0.125 & 0.281 & 0.346 & 0.394 & 0.430 & 6.615\\ \hline
%		\textbf{\fbmusic} & 0.714 & 0.893 & 0.935 & 0.955 & 0.966 & 7.113\\ \hline
		\textbf{\fbmovies} & 0.821 & 0.945 & 0.970 & 0.980 & 0.984 & 12.856\\ \hline
%		\textbf{\fbbooks} & 0.315 & 0.516 & 0.605 & 0.682 & 0.745 & 3.133\\ \hline
	\end{tabular}
	\caption{1-Robustness for different values of M. Column F.A. denotes the Feature Average number per item.}
	\label{tbl:robustness}
\end{table}
In this case, we focus on the CS setting which provides the best results in terms of accuracy.  
For a better understanding of the obtained results, we start by focusing on \yahoo which apparently has bad behavior. Table \ref{tbl:meaningfulness} showed that \framework was able to guess 10 on 12 different features for \yahoo. In this experiment, we remove one of the ten features (thus, based on Table \ref{tbl:meaningfulness}, \framework will guess an average of $10 - 1 = 9$ features). Since the number of features is 12 we have 3 remaining "slots". What we measure now is if \framework is able to guess the removed feature in these "slots". Results in Table \ref{tbl:robustness} show that our method is able to put the removed feature in one of the three slots the 48.7\% of the times starting from 747 overall features.
This example should help the reader to appreciate even more \texttt{Facebook Movies} results.  
Hence, we could confidently assess that \framework is able to propose meaningful features as we asked with RQ3.

\section{Related Work}\label{sec:related}
In recent years, several interpretable recommendation models that exploit matrix factorization have been proposed. It is well-known that one of the main issues of matrix factorization methods is that they are not easily interpretable (since latent factors meaning is basically unknown). One of the first attempts to overcome this problem was proposed in \cite{DBLP:conf/sigir/ZhangL0ZLM14}. 
%\cite{DBLP:journals/corr/abs-1708-06409}
In this work, the authors propose Explicit Factor Model (EFM). Products' features and users' opinions are extracted with phrase-level sentiment analysis from users' reviews to feed a matrix factorization framework. After that, a few improvements to EFM have been proposed to deal with temporal dynamics \cite{DBLP:conf/www/Zhang0ZLLZM15} and to use tensor factorization \cite{DBLP:conf/sigir/ChenQZX16}. In particular, in the latter the aim is to predict both user preferences on features (extracted from textual reviews) and items. This is achieved by exploiting the Bayesian Personalized Ranking (BPR) criterion \cite{DBLP:conf/uai/RendleFGS09}. 
%Eventually, these preferences are combined to produce recommendation lists. We considered this work interesting because they also adopt a pair-wise learning to rank algorithm, but this is really different from ours since we exploited BPR to explicitly train the feature vectors to rank items. 
Further advances in MF-based interpretable recommendation models have been proposed with Explainable Matrix Factorization (EMF) \cite{DBLP:conf/www/AbdollahiN16} in which the generated explanations are based on a neighborhood model. 
%We differ from EMF as they do not take advantage of any external data source, other than they use a completely different model. Moreover, they introduced an additional regularizer into the factorization model to constrain users' vectors training. 
Similarly, in \cite{DBLP:journals/corr/AbdollahiN16} an interpretable Restricted Boltzmann Machine model has been proposed. It learns a network model (with an additional visible layer) that takes into account a degree of explainability. 
%To measure it, they  defined an ad-hoc \textit{Explainability Score}. 
Finally, an interesting work incorporates sentiments and ratings into a matrix factorization model, named Sentiment Utility Logistic Model (SULM) \cite{DBLP:conf/kdd/Bauman0T17}.
%, generates a novel kind of explanations composed by both items and features. 
In \cite{DBLP:conf/recsys/RanaB17} recommendations are computed by generating and ranking  personalized explanations in the form of explanation chains.
%\cite{vlachos2018addressing, DBLP:conf/icde/HeckelVPD17, DBLP:journals/ibmrd/VlachosVHL16}
OCuLaR \cite{vlachos2018addressing} provides interpretable recommendations from positive examples based on the detection of co-clusters between users (clients) and items (products). 
%The recommendation comes with the corresponding user-item co-clusters, which provide much more detailed information than usual collaborative-based explanations.
In \cite{DBLP:conf/ijcai/HuJCC18} authors propose a Multi Level Attraction Model (MLAM) in which they build two attraction models, for cast and story. 
%Moreover, the story model is built upon two attraction models: for story level and for sentence level. 
The interpretability of the model is then provided in terms of attractiveness of Sentence level, Word level, and Cast member.
In \cite{DBLP:conf/kdd/PeakeW18} the authors train a matrix factorization model to 
%complete the $U \times I$ matrix. They then use the complete (approximated) ratings matrix to 
compute a set of association rules that interprets the obtained recommendations.
In \cite{DBLP:journals/corr/DhurandharOP16} the authors prove that, given the conversion probabilities for all actions of customer features, it is possible to transform the original historical data to a new space in order to compute a set of interpretable recommendation rules.
%The design of a new explainable model is useless if it does not match with any existing explanations generation technique. For this reason we deepened the different ways of generating feature-based explanations in order to provide a flexible but accurate model.
%Probably the most used style of explanations of this kind are the content-based ones \cite{symeonidis2008justified}. In their most simple form, authors consider similarities between items by taking into account both item properties and user ratings. Among the works that exploit content information to produce explainable recommendations, Tagsplanations \cite{DBLP:conf/iui/VigSR09} is worth to mention. It is fed by community tags and it exploits a relevance measure to weight tags w.r.t. items and user preferences. Furthermore, also demographic-based recommendations explanations have been inspected \cite{DBLP:journals/kais/ZhaoLHWWL16}, in order to recommend items for particular types (age, location, gender) of users. 
The core of our model is a general Factorization Machines (FM) model \cite{DBLP:conf/icdm/Rendle10}.
Nowadays FMs are the most widely used factorization models because they offer a number of advantages w.r.t. other latent factors models such as SVD++ \cite{DBLP:conf/kdd/Koren08}, PITF \cite{DBLP:conf/wsdm/RendleS10}, FPMC \cite{DBLP:conf/www/RendleFS10}. First of all, FMs are designed for a generic prediction task while the others can be exploited only for specific tasks. Moreover, it is a linear model and parameters can be estimated accurately even in high data sparsity scenarios. Nevertheless, several improvements have been proposed for FMs. For instance Neural Factorization Machines \cite{DBLP:conf/sigir/0001C17} have been developed to fix the inability of classical FMs to capture non linear structure of real-world data. This goal is achieved by exploiting the non linearity of neural networks. Furthermore, Attentional Factorization Machines \cite{DBLP:conf/ijcai/XiaoY0ZWC17} have been proposed that use an attention network to learn the importance of feature interactions. Finally, FMs have been specialized to better work as Context-Aware recommender systems \cite{DBLP:conf/sigir/RendleGFS11}. 
\section{Conclusion and Future Work}\label{sec:conclusion}
In this work, we have proposed an interpretable method for recommendation scenario, \framework, in which we bind the meaning of latent factors for a Factorization machine to data coming from a knowledge graph. We evaluated \framework on two different publicly available datasets on different sets of semantics-aware features. In particular, in this paper we considered Ontological, Categorical and Factual information coming from \dbpedia and we have shown that the generated recommendation lists are more precise and personalized. Summing up, performed experiments show that: (RQ1) the learned model shows very good performance in terms of accuracy and, at the same time, is effectively interpretable; (RQ2) the computed features are semantically meaningful; (RQ3) the model is robust regarding computed features. 
In the future we want to test the \framework performance in different scenarios, other than recommender systems. Moreover, the model can be improved in many different ways. 
Different relevance metrics could be beneficial in different scenarios, as the method itself is agnostic to the specific adopted measure. This work focused on the items' vector; however, an interesting key point would be analyzing the learned users' vectors to extract more accurate profiles.  Furthermore, it would be useful to exploit \framework in order to provide suggestions to knowledge graphs maintainers while adding relevant missing features to the knowledge base. In this direction, we would like to evaluate our approach in knowledge graph completion task. 
 \bibliographystyle{splncs04}
 \bibliography{bibliographywa}
\end{document}